\documentclass[times,twocolumn,final,authoryear]{elsarticle}

\usepackage{jasr}
\usepackage{framed,multirow}

\usepackage{amssymb}
\usepackage{latexsym}

\usepackage{url}
\usepackage{xcolor}
\definecolor{newcolor}{rgb}{.8,.349,.1}
\usepackage[citebordercolor=white]{hyperref}

\usepackage{hyperref}
\hypersetup{
    colorlinks=true,
    linkcolor=blue,
    filecolor=blue,      
    urlcolor=blue,
    citecolor=blue,
    filecolor=blue,
}
\def\1{\c{c}}
\def\2{\c{C}}
\def\3{\.{I}}
\def\4{\"{a}}
\def\5{{\i}}
\def\6{$\beta$}
\def\7{\"{o}}
\def\8{\"{O}}
\def\9{\c{s}}
\def\0{\c{S}}
\def\*{\"{u}}
\def\?{\"{U}}
\def\;{\u{g}}
\def\:{\u{G}}
\usepackage{wrapfig}
\usepackage{float}
\usepackage{inputenc}

\journal{Advances in Space Research}

\begin{document}

\verso{T. Ergin \textit{etal}}
\begin{frontmatter}

\title{Origin of the gamma-ray emission from supernova remnant HB9}

\author[1]{T\*l\*n Ergin\corref{cor1}\fnref{fn1}}
\cortext[cor1]{Corresponding author}
\fntext[fn1]{Now at Middle East Technical University, Physics Group, Northern Cyprus Campus, 99738 Kalkanli via Mersin 10, Turkey.}
\ead{ergin.tulun@gmail.com}
\author[2]{Lab Saha\corref{cor1}\fnref{fn2}}
\ead{lab.saha@gmail.com}
\fntext[fn2]{Now at Harvard-Smithsonian Center for Astrophysics, Cambridge, MA.}
\author[3]{Hidetoshi Sano \corref{cor1}}
\ead{hidetoshi.sano@nao.ac.jp}
\author[4]{Aytap Sezer}
\author[5]{Ryo Yamazaki}
\author[6]{Pratik Majumdar}
\author[7]{Yasuo Fukui}

\address[1]{TUBITAK Space Technologies Research Institute, ODTU Campus, 06531, Ankara, Turkey}
\address[2]{PARCOS and Department of EMFTEL, Universidad Complutense de Madrid, E-28040 Madrid, Spain}
\address[3]{National Astronomical Observatory of Japan, Mitaka, Tokyo 181-8588, Japan}
\address[4]{Department of Electrical-Electronics Engineering, Avrasya University, 61250, Trabzon, Turkey}
\address[5]{Department of Physics and Mathematics, Aoyama Gakuin University, 5-10-1 Fuchinobe, Sagamihara 252-5258, Japan}
\address[6]{Saha Institute of Nuclear Physics, HBNI, 1/AF Bidhannagar, Kolkata 700064, India}
\address[7]{Department of Physics, Nagoya University, Furo-cho, Chikusa-ku, Nagoya 464-8601, Japan}

\received{.. May 2021}
\finalform{.. May 2021}
\accepted{.. May 2021}
\availableonline{.. May 2021}
\communicated{S. Sarkar}

\begin{abstract}
HB9 (G160.9+2.6) is a mixed-morphology Galactic supernova remnant (SNR) at a distance of $\sim$0.6 kpc. Previous analyses revealed recombining plasma emission in X-rays and an expanding shell structure in H\,{\sc i} and CO emission, which were correlating with the spatial extent of HB9. In GeV energies, HB9 was found to show extended gamma-ray emission with a morphology that is consistent with the radio continuum emission showing a log-parabola-type spectrum. The overlap reported between the gas data and the excess gamma-ray emission at the southern region of the SNR's shell could indicate a possible interaction between them. We searched for hadronic gamma-ray emission signature in the spectrum to uncover possible interaction between the molecular environment and the SNR. Here we report the results of the gamma-ray spectral modelling studies of HB9. 
\end{abstract}


\end{frontmatter}

\section{Introduction}
\label{intro}
Detection of gamma-ray emission from supernova remnants (SNRs) and their neighbourhood is key for understanding the origin, acceleration, and propagation of cosmic rays (CRs) with very high energies up to 10$^{15}$ eV.  

 CRs (i.e.$~\!$protons) that escape the shock front of SNRs enter the surrounding ambient interstellar medium (ISM). Here they interact with protons in the ambient gas or molecular clouds (MCs) through the {\it hadronic} interactions, which create neutral pions that later decay into gamma rays \citep{Ah96, Ga09, Fu09}. This hadronic interaction model is called the {\it illuminated cloud model} or {\it runaway CR model} \citep{Pu03, Pu05, Oh10, Li10, El11, Ma13, Ce19}. 
 
 Therefore, gamma rays from nearby MCs are crucial to probe the highest energy protons escaping SNRs. The second well-known hadronic model is the {\it crushed cloud model} or {\it direct interaction model}, where the SNR's shock wave interacts with MCs \citep{By00, Uc10, Oh11, Ta15, Le15, Ca16}. This latter model is usually observed in middle-aged SNRs.  

In the leptonic models \citep{Ah99, Po06}, accelerated electrons produce gamma rays through the inverse-Compton (IC) and Bremsstrahlung processes. This type of gamma-ray emission is usually accompanied by synchrotron emission observed in radio wavebands. IC emission is produced by the interaction of the accelerated electrons with the ambient photons, like cosmic microwave background (CMB) emission. Highly energetic electrons may be accelerated at the shock fronts of SNRs or they may originate from pulsars. 

HB9 (G160.9+2.6) is a middle-aged mixed-morphology (MM) supernova remnant (SNR) characterised by centre-filled thermal X-rays and a radio shell \citep{RhPe98}.

HB9 has a large angular size and shell-like morphology with a non-thermal radio spectral index of $\sim$0.64 (e.g. \citealt{LeRo91, LeAs95, LeTi07}).

The kinematic distance to HB9 was calculated to be 0.6 $\pm$ 0.3 kpc, which was derived from the expanding shell structure around the remnant \citep{Se19} using H\,{\sc i} and $^{12}$CO($J$ = 1$-$0) data in the velocity range of $-$10.5 km s$^{-1}$ and $+$1.8 km s$^{-1}$. More recently, \citet{Zh20} estimated the extinction distances to 33 SNRs. Using {\it Gaia} stars on the line of sight to HB9 with infrared/optical photometry, they obtained extinction of each star as a function of the distance determined by {\it Gaia}. Their estimated distance is 0.54 $\pm$ 0.10 kpc, which is fully consistent with results of \citet{Se19} (0.6 $\pm$ 0.3 kpc) that was calculated using $^{12}$CO($J$ = 1$-$0) data. This result clearly shows the association between HB9 and interstellar gas at the distance of $\sim$0.6 kpc.

It was also reported that HB9 contained recombining (over-ionised) plasma (RP) in {\it Suzaku} spectra. More recently, \citet{Sai20} re-analysed {\it Suzaku} data to understand the nature of the hard X-ray component reported by previous X-ray study \citep{Ya93}. They found no significant hard X-ray emission from the central region and no significant feature of an over-ionised plasma in HB9. Their spectra extracted from the centre of HB9 are well-fitted with a model consisting of a collisional ionisation equilibrium (CIE) and ionising plasma (IP).
\begin{figure*}
\centering \vspace*{1pt}
\includegraphics[width=1.0\textwidth]{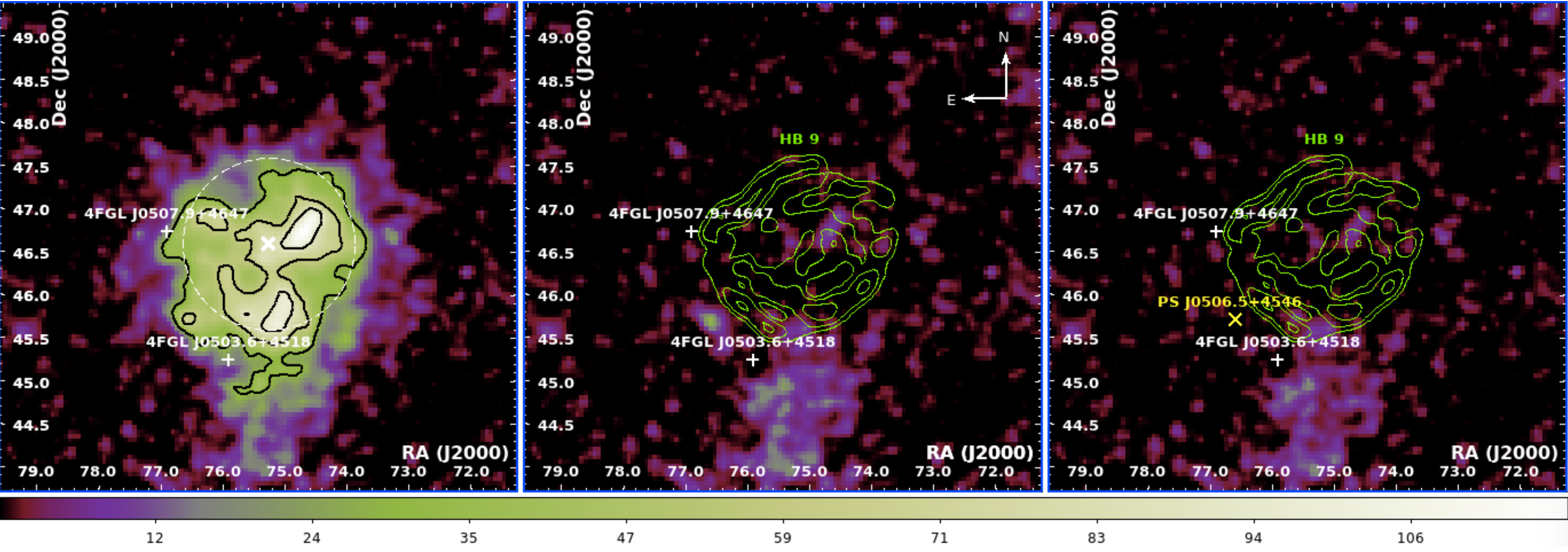}
\vspace{-0.8cm}
\caption{Gamma-ray TS maps shown in the energy range of 1$-$300 GeV. Left Panel: HB9 was not added into the gamma-ray background model while producing the TS map. Black contours show the TS levels of 30, 59, 89, 119. The white cross and the dashed circle show the radio centre and extension of HB9; Middle Panel:  The 4850 MHz radio continuum map of HB9 is used as a template in the gamma-ray background model (shown with green contours); Right Panel: In addition to the 4850 MHz radio continuum template of HB9, a point-source (PS J0506.5+4546) was added to the gamma-ray background model. The position of PS J0506.5+4546 is shown with a yellow cross. For further information see \citet{Se19}. There is an excess of diffuse and extended gamma-ray emission observable towards the south of the radio-shell of HB9.}
\label{figure1}
\end{figure*}
\begin{figure*}
\centering \vspace*{1pt}
\includegraphics[width=0.7\textwidth]{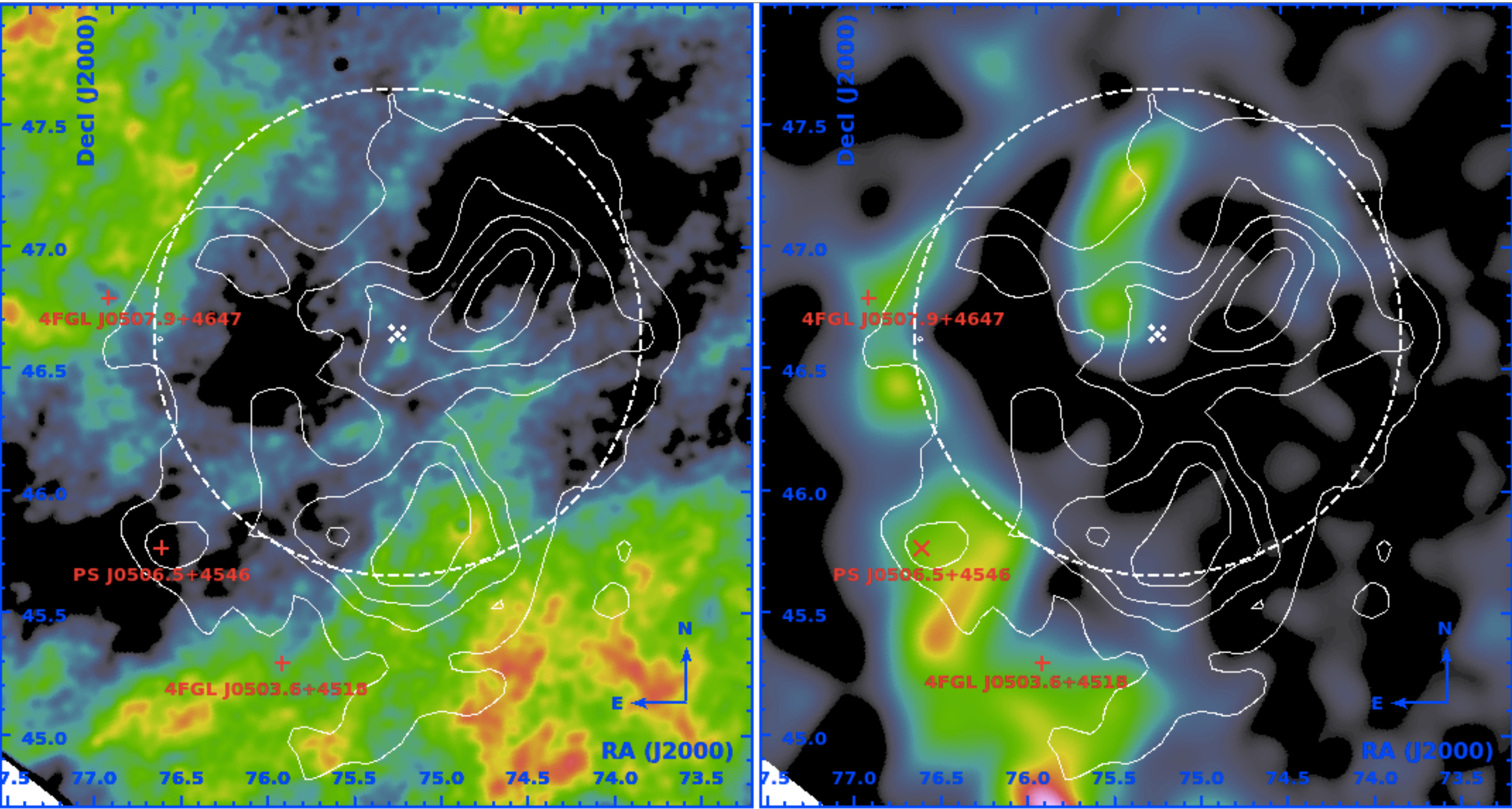}
\vspace{-0.2cm}
\caption{Intensity distributions of H\,{\sc i} (left panel) and $^{12}$CO($J$ = 1--0) (right panel) in the gamma-ray analysis region are shown in Equatorial coordinates. The integration velocity of H\,{\sc i} and CO is from $-10.5$ km s$^{-1}$ to $+1.8$ km s$^{-1}$. The superposed white contours correspond to TS values of 25, 49, 64, 81, 100 in the energy range of 0.2$-$300 GeV. The white cross and the dashed circle show the radio centre and extension of HB9.}
\label{figure2}
\end{figure*}

In the GeV gamma-ray energy range, many middle-aged MM SNRs were detected by the Large Area Telescope (LAT) on board the Fermi Gamma-ray Space Telescope ({\it Fermi}) (e.g. \citealt {Ab09, Ab10a, Ab10b, CaSl10, Er17}). HB9 has recently been found to be one of these MM SNRs (\citealt{Ar14}, \citealt{Se19}). Using 5.5 years of {\it Fermi}-LAT data, \citet{Ar14} detected extended gamma-ray emission centred at the position of HB9 with a significance of 16$\sigma$ above 0.2 GeV. 10 years of {\it Fermi}-LAT data was analysed and investigated the gamma-ray emission morphology and spectrum within the energy range of 0.2$-$300 GeV. HB9 was detected as an extended gamma-ray source with a significance of $\sim$25$\sigma$. The best-fitted spatial model to HB9 in the 1$-$300 GeV energy range was the 4850 MHz radio continuum map (taken by the Green Bank Telescope), that has an approximate size of 2$^{\circ}$. In addition, a new point-source (PS J0506.5+4546) was marginally detected ($\sim$6$\sigma$) right outside the extended region of HB9.

The spectrum of HB9 was best-fitted to a log-parabola (LP) function with indices of $\alpha$ = 2.36 $\pm$ 0.05 and $\beta$ = 0.14 $\pm$ 0.05. The total energy flux was found to be (1.51 $\pm$ 0.08) $\times$ 10$^{-5}$ MeV cm$^{-2}$ s$^{-1}$ in the energy range of 0.2$-$300 GeV. These results were consistent with previous studies \citep{Ar14}. In the TeV energy range MAGIC observations revealed no detection with an integral flux upper limit of 1.6 $\times$ 10$^{-11}$ cm$^{-2}$ s$^{-1}$ for energies above 270 GeV \citep{Ca09}.  

This work contains a short summary of our previous paper, \citet{Se19}, and an addition of spectral modelling to it. Here, we aim to investigate possible gamma-ray emission mechanisms (leptonic/hadronic) by modelling the gamma-ray spectrum obtained from the whole SNR HB9, which was reported in the energy range of 0.2$-$300 GeV. The paper is organised as follows: In Section 2, we summarise gamma-ray data analysis results and we estimate the mean ISM proton density for the remnant (Section 2.1). We then report our preliminary results about the SED modelling of HB9 in Section 2.2, and we close this work by conclusions in Section 3.

\section{Analysis and Results}
\label{ana_res}
Figure \ref{figure1} shows the distribution of the GeV gamma-rays based on the TS maps produced in the energy range of 1 - 300 GeV, where the left panel shows the TS map without including the radio continuum template of HB9 and the newly discovered gamma-ray point source (PS J0506.5+4546) in the gamma-ray background model. The right panel shows the TS morphology where both PS J0506.5+4546 and the radio continuum template are included in the background model. Two sources from the Fourth {\it Fermi}-LAT sources (4FGL) catalog \citep{Fermi20} are shown by white markers.

PS J0506.5$+$4546 was detected at a significance level of about 6 sigma by applying the best-fit spatial model (i.e. Radio Template $+$ 1 P. S.), but we assumed PS J0506.5$+$4546 to be physically not related to HB9, based on the assumption that it lies outside the shell of the SNR.

However, there are currently no known counterparts corresponding to this source, apart from the  positional overlap with the gas/MCs data, which we have shown. So, it is not clear whether PS J0506.5+4546 is another source independent of HB9 or it is part of the SNR’s shell interacting with the ambient gas/MCs.

There is another region of diffuse gamma-ray emission towards even more south of the SNR, but neither in our previous paper, nor in this study we analysed this “rather large/extended region, because we assumed that this region might not be directly related to HB9, mainly because it lies outside the radio emission region of the SNR. In addition, this large and diffuse gamma-ray emission region might be an extension of 4FGL 0503.6+4518, which we also assumed to be not related to the SNR (but included as a point-source into our gamma-ray background model).

\subsection{Estimation of the Mean ISM Proton Density for SNR HB9}
\label{est_dens}
In Figure \ref{figure2}, the distribution of the H\,{\sc i} (left panel) and $^{12}$CO($J$ = 1$-$0) (right panel) data is shown in the analysis region of HB9, where the 408 MHz radio continuum intensity contours are overlaid. We calculated a mean ISM proton density of $\sim$60 cm$^{-3}$ by assuming following parameters as given by  \citet{Sa19a}:  
\vspace{0.2cm}
\begin{itemize}
\item Distance: 0.6 kpc,
\item Shell radius: 11 pc ($\sim$1.05 deg),
\item Shell thickness: 3.5 pc ($\sim$0.33 deg),
\item CO-to-H$_2$ conversion factor: 2.0 $\times$ 10$^{20}$ cm$^{-2}$ (K km s$^{-2}$)$^{-1}$,
\item H\,{\sc i}: Optically thin assumption.
\end{itemize}
The total column density is taken as the summation of twice the molecular hydrogen column density and the proton column density for atomic hydrogen.  

\subsection{SED Modelling of SNR HB9 and Preliminary Results}
\label{sed_mod}
We modelled the SED of HB9, which was reported by \citep{Se19}. We have considered both leptonic, i.e. bremsstrahlung and inverse Compton (IC) scattering, and hadronic models to explain the observed data \citep{Bl70,Ke06}. For the relativistic cosmic-ray electrons and protons we assumed an energy spectrum of the form 

\begin{equation}\label{eq:01}
\frac{dN}{dE} = A  E^{-\alpha '} e^{(-E/E_{\rm cut})},
\nonumber
\end{equation}
where $\alpha '$ is the spectral index, {\it A} is the amplitude, and $E_{\rm cut}$ is the cut-off energy of the particle spectrum. 

The expression for the total energy (E$_{total}$) of particles is  
\begin{equation}\label{eq:02}
    E_{total} = \int_{E_1}^{E_2} E {dN\over dE} dE, \nonumber
\end{equation}
where $E_1$ and $E_2$ are limits of the energy of particles. 

\begin{figure*}
\centering
\includegraphics[width=0.6\textwidth]{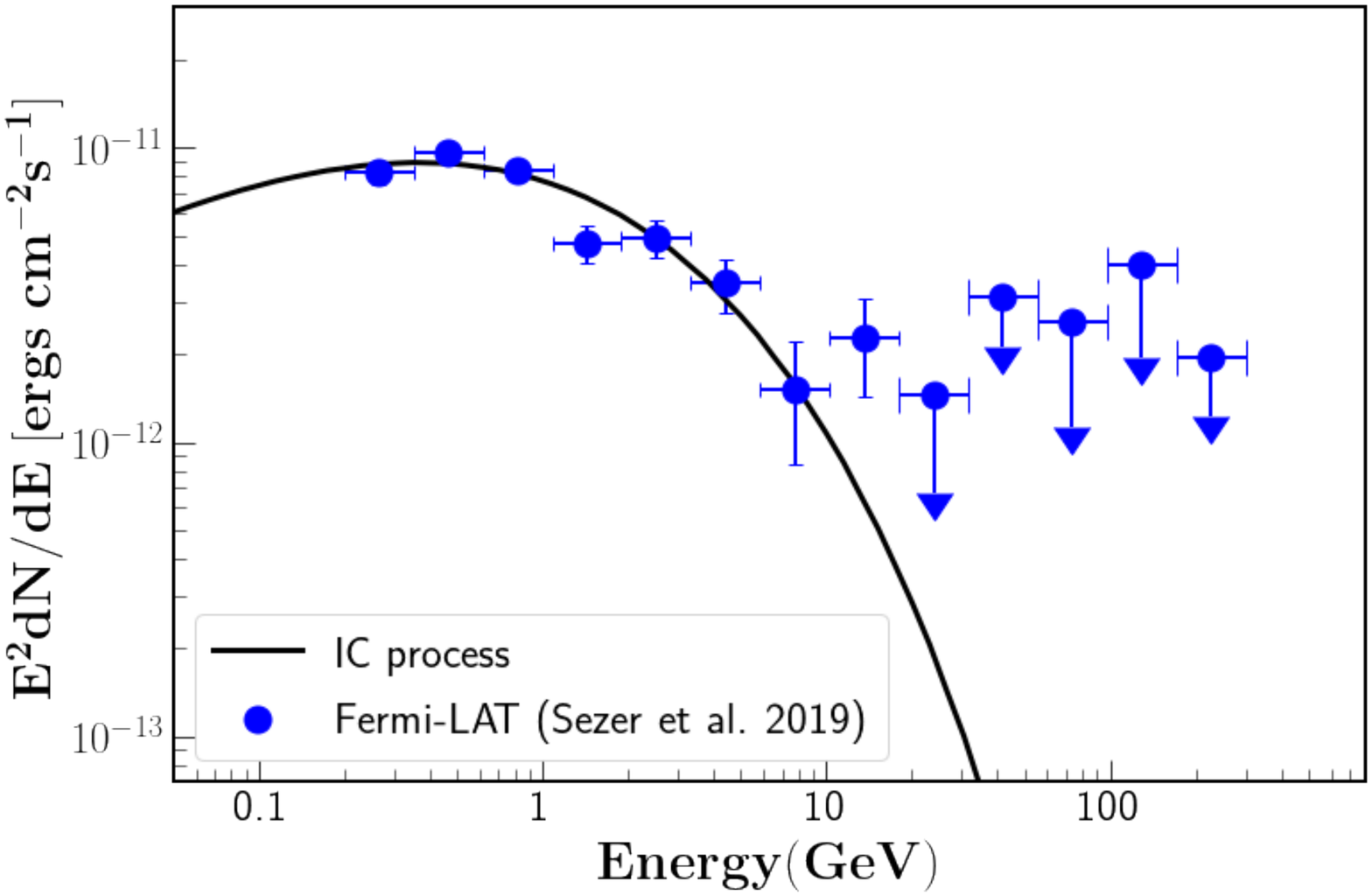}
\includegraphics[width=0.6\textwidth]{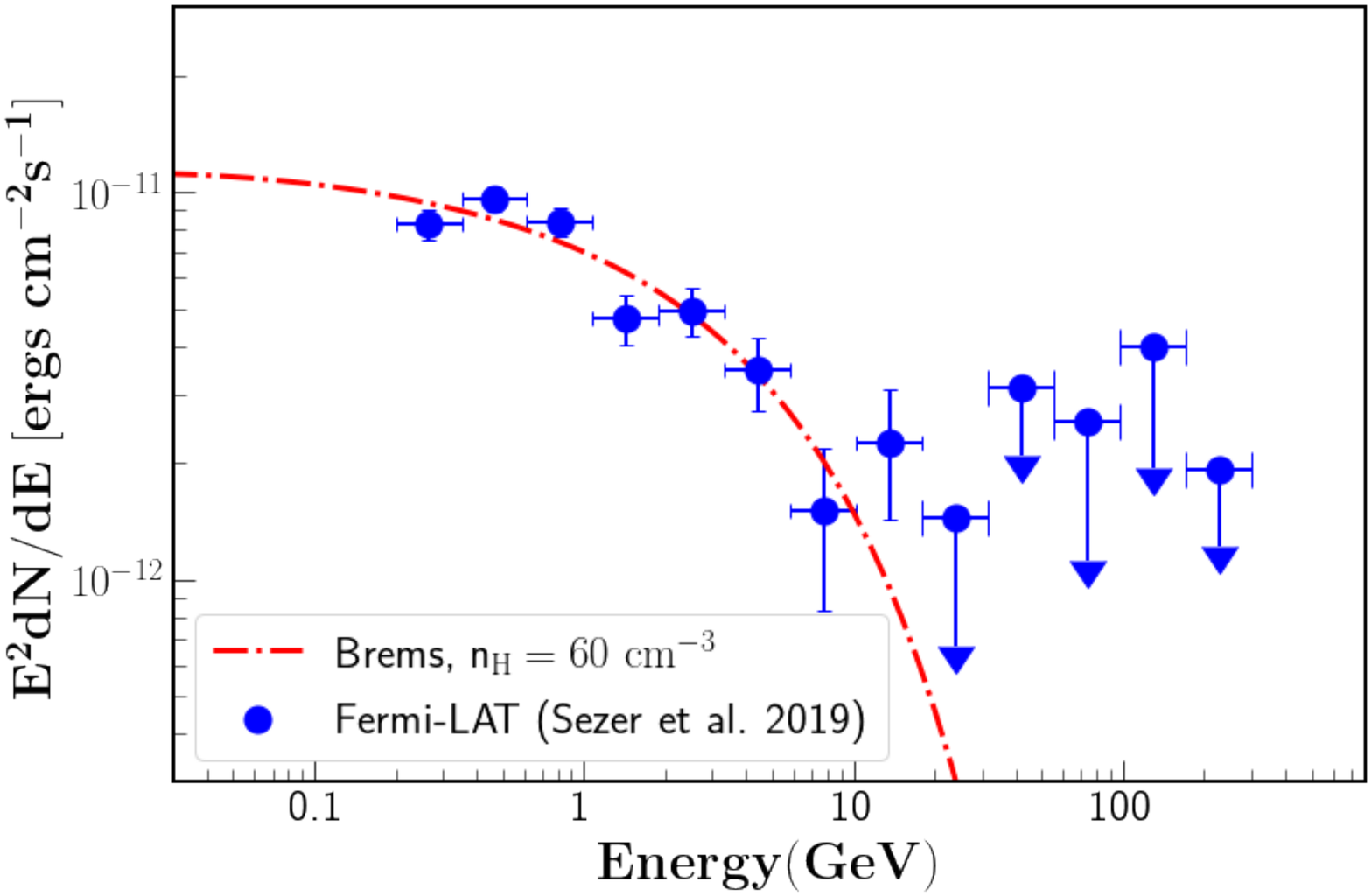}
\includegraphics[width=0.6\textwidth]{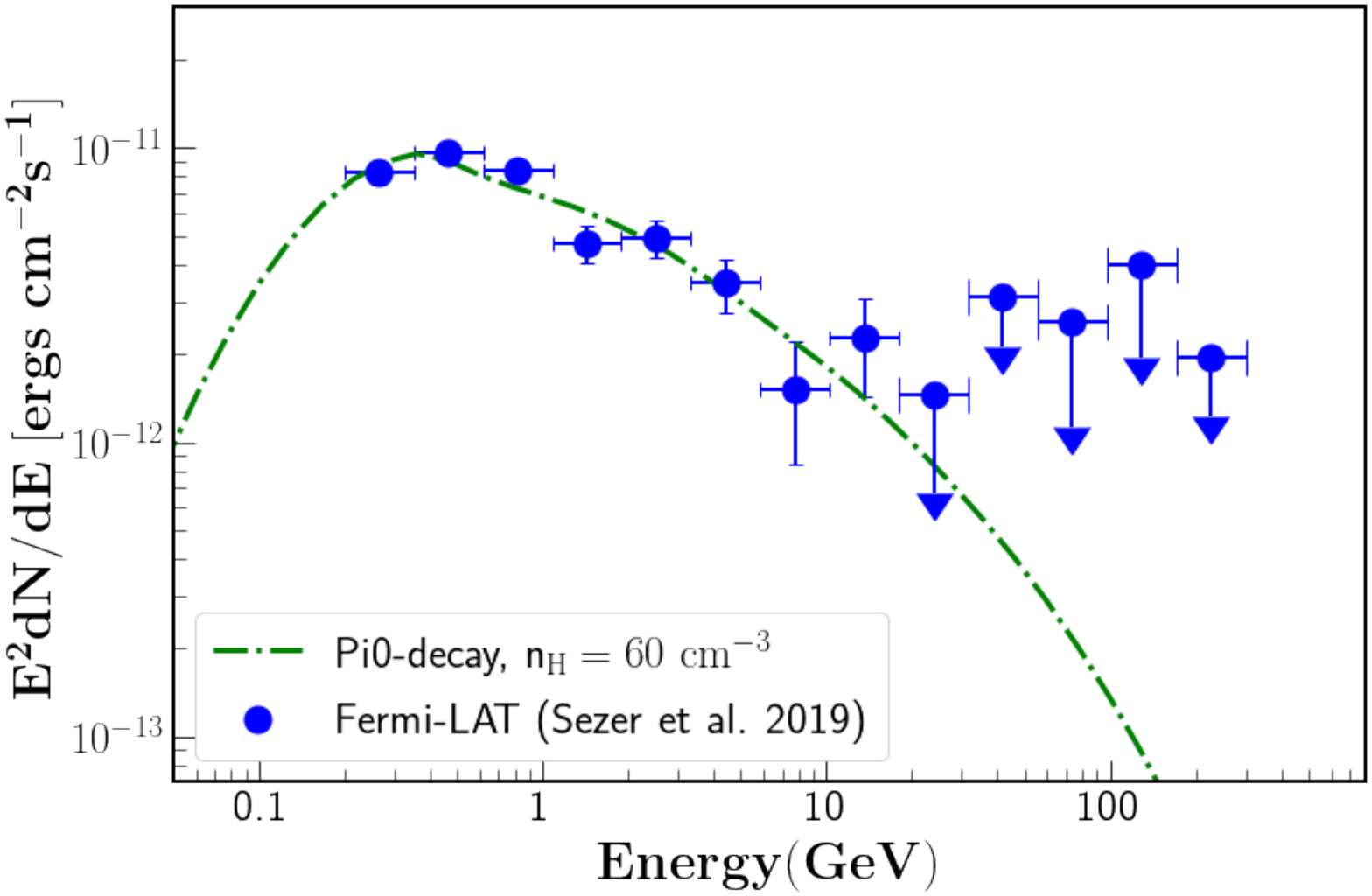}
\vspace{-0.2cm}
\caption{Gamma-ray SED and the preliminary leptonic models (IC-scattering on the upper plot and Bremsstrahlung on the middle plot) and the hadronic (on the bottom plot) model fitted by assuming n$_{H}$ = 60 cm$^{-3}$.}
\label{figure3}
\end{figure*}

We used CMB and interstellar dust emission for the calculation of IC fluxes \citep{Ma83}.  
\begin{table}
\begin{minipage}{90mm}
\begin{center}
\caption{The estimated spectral parameters for the electron and proton spectra. The total energy (E$_{total}$) calculated for particles for the hadronic ($\pi^{0}$-decay) and leptonic (Bremsstrahlung and IC scattering) spectral models fitted to the gamma-ray SED of HB9 assumes a distance of $\sim$0.6 kpc and n$_{\rm H}$ = 60 cm$^{-3}$.} 
\vspace{0.2cm}
\begin{tabular}{@{}p{1.5cm}p{2.0cm}p{1.4cm}p{1.9cm}@{}}
 \hline
$~$                 & Bremsstrahlung &$\pi^{0}$-decay  & IC-scattering      \\ 
\hline
$\alpha '$          & 2.21           & 2.69           & 1.74  \\ 
E$_{\rm cut}$ [GeV]	& 10             & 1000           & 200  \\ 
E$_{\rm total}$  [erg] & 1.45$\times$10$^{47}$ & 2.25$\times$10$^{47}$  & 1.90$\times$10$^{48}$   \\ 
\hline
\label{table1}
\end{tabular}
\end{center}
\end{minipage}
\vspace{-0.5cm}
\end{table}
For bremsstrahlung and pion-decay processes, we used the mean ISM proton density (n$_{\rm H}$) to be 60 cm$^{-3}$ and the proton-proton cross section from \citet{Ke06} was used for the pion-decay modelling. The preliminary fitted spectral models are shown in Figure \ref{figure3}. The total energy of particles for each of the different fitted spectral models is given in Table \ref{table1}.

\section{Discussion and Conclusion}
\label{conc}
\vspace{-0.3cm}
We modelled the GeV gamma-ray emission spectrum of HB9, which was obtained from our previous analysis reported in S19. Using H\,{\sc i} and $^{12}$CO($J$ = 1$-$0) data, we estimated the ISM proton density of $\sim$60 cm$^{-3}$, which is an average value calculated by assuming the shell radius and thickness to be 11 pc and 3.5 kpc, respectively.  

Although the classical signs of shock-cloud interaction (e.g., CO line-broadening, the high kinetic temperature of shocked clouds) are not present for HB9, an expanding CO/HI shell related to HB9 was found (shown in Figure 4 Panel (d) of \citep{Se19}). In general, the expanding shell is thought to be formed by strong stellar winds from the progenitor of the SNR and the SNR shock is now likely interacting with the wind-wall (e.g. \citealt{Sa17, Sa19b}), which are walls of the stellar wind bubble formed in the ISM by swept up cold and warm gas, in neutral and ionised state \citep{Bo07}. Therefore, accelerated CRs by the shock fronts of HB9 are mainly interacting with the inner wall of the stellar wind bubble. The integration velocity range, therefore, can be determined quite uniquely. We confirmed that the density contribution is $\sim$70\% for atomic hydrogen and $\sim$30\% for molecular hydrogen. 

By looking at the gas distribution overlapping HB9, a dominating hadronic gamma-ray emission was expected as a result of our SED modelling. However, the observed gamma-ray spectrum of HB9 can be explained by both leptonic and hadronic models. Within the context of the leptonic model, we found that both IC and Bremsstrahlung can explain the observed SED individually for two different sets of model parameters. 

The reason for this result is mainly due to the assumption that ISM to be uniformly distributed within the shell, with a density of $\sim$60 cm$^{-3}$, which is not the case. The gaseous interstellar material overlaps only at certain regions of the SNR's shell, which has to be taken into account while doing spectral analysis and modelling the SED. So, if we consider both IC and Bremsstrahlung processes together to explain the observed SED for a set of model parameters, we need a robust estimate of the ambient matter density which is currently lacking.

Although the overall fitting of the models to the gamma-ray SED, does not show preference of any model over the other (i.e. Leptonic/Hadronic), we note that the SED of HB9 might give some clues about the dominating gamma-ray emission model. Namely, when we compare our spectral modelling results for HB9 with results from \citet{Ac13}, we see that interacting SNRs, such as IC 443 and W44, present a spectrum that fits better to a hadronic gamma-ray emission model. In particular, at gamma-ray energies lower than 200 MeV, there is a clear separation between Bremsstrahlung models and the Pi-zero Decay model, where  the flux levels of the latter one decrease towards lower energies. 

For the gamma-ray SED of HB9 at energies below 200 MeV, although results of spectral modelling hint a dominating hadronic emission which is backed up by CO and HI intensity maps in Figure \ref{figure2}, more data and detailed analyses are required to conclude whether the hadronic model is really dominating for HB9. 

\section{Acknowledgments}
\vspace{-0.2cm}
We would like to thank our referees at the {\it Advances in Space Research} journal for their valuable comments.
\vspace{0.2cm}
$~$\\
{\it {\bf Facilities}}: {\it Fermi}-LAT, Dominion Radio Astrophysical Observatory, Harvard-Smithsonian Center for Astrophysics 1.2 m MMW-radio Telescope, Green Bank Telescope.\\



\vspace{-0.5cm}
\begin{thebibliography}{}
\bibitem[\protect\citeauthoryear{Abdo et al.}{2009}]{Ab09} Abdo, A. A., Ackermann, M., Ajello, M. et al., 2009. Fermi LAT Discovery of Extended Gamma-Ray Emission in the Direction of Supernova Remnant W51C. ApJL 706, L1-L6. https://doi.org/10.1088/0004-637X/706/1/L1.

\bibitem[\protect\citeauthoryear{Abdo et al.}{2010a}]{Ab10a} Abdo, A. A. et al., 2010a. Observation of Supernova Remnant IC 443 with the Fermi Large Area Telescope. ApJ 712, 459-468. https://doi.org/10.1088/0004-637X/712/1/459.

\bibitem[\protect\citeauthoryear{Abdo et al.}{2010b}]{Ab10b} Abdo, A. A. et al., 2010b. Fermi Large Area Telescope Observations of the Supernova Remnant W28 (G6.4-0.1). ApJ 718, 348-356. https://doi.org/10.1088/0004-637X/718/1/348.

\bibitem[\protect\citeauthoryear{Ackermann et al.}{2013}]{Ac13} Ackermann, M. et al. (The Fermi-LAT collaboration), 2013. Detection of the Characteristic Pion-Decay Signature in Supernova Remnants. Science 339, 807-811.  https://doi.org/10.1126/science.1231160.

\bibitem[\protect\citeauthoryear{Aharonian \& Atoyan}{1996}]{Ah96}Aharonian, F. A., Atoyan, A. M., 1996. On the emissivity of $\pi^{0}$-decay gamma radiation in the vicinity of accelerators of galactic cosmic rays. A\&A 309, 917-928.

\bibitem[\protect\citeauthoryear{Aharonian \& Atoyan }{1999}]{Ah99}Aharonian, F. A., Atoyan, A. M., 1999. On the origin of TeV radiation of SN 1006. A\&A 351, 330-340.

\bibitem[\protect\citeauthoryear{Araya}{2014}]{Ar14} Araya, M., 2014. Fermi LAT observation of supernova remnant HB9. MNRAS 444, 860-865. https://doi.org/10.1093/mnras/stu1484.

\bibitem[\protect\citeauthoryear{Blumenthal \& Gould}{1970}]{Bl70} Blumenthal, G.R., Gould, R.J., 1970. Bremsstrahlung, synchrotron radiation, and Compton scattering of high-energy electrons traversing dilute gases. Rev. Mod. Phys. 42, 237–271. http://dx.doi.org/10.1103/RevModPhys.42.237

\bibitem[\protect\citeauthoryear{R. Boomsma}{2007}]{Bo07} Boomsma, R. 2007. The disk-halo connection in NGC 6946 and NGC 253. Ph.D. Thesis, University of Groningen, s.n., ISBNs: 9789036729468.

\bibitem[\protect\citeauthoryear{Bykov et al.}{2000}]{By00} Bykov, A. M., Chevalier, R. A., Ellison, D. C., \& Uvarov Y. A., 2000. Nonthermal Emission from a Supernova Remnant in a Molecular Cloud. ApJ 538, 203-216. https://doi.org/10.1086/309103.

\bibitem[\protect\citeauthoryear{Cardillo et al.}{2016}]{Ca16} Cardillo, M., Amato, E., \& Blasi P., 2016. Supernova remnant W44: a case of cosmic-ray reacceleration. A\&A 595, A58.
https://doi.org/10.1051/0004-6361/201628669.

\bibitem[\protect\citeauthoryear{Carmona}{2009}]{Ca09} Carmona, E., Costado, M. T., Font, L., Zapatero, J., 2009. Observation of selected SNRs with the MAGIC Cherenkov Telescope. arXiv:0907.1009.

\bibitem[\protect\citeauthoryear{Castro \& Slane}{2010}]{CaSl10} Castro, D., Slane, P., 2010. Fermi Large Area Telescope Observations of Supernova Remnants Interacting with Molecular Clouds. ApJ 717, 372-378.
https://doi.org/10.1088/0004-637X/717/1/372. 

\bibitem[\protect\citeauthoryear{Celli et al.}{2019}]{Ce19} Celli, S., Morlino, G., Gabici, S., \& Aharonian, F. A., 2019. Exploring particle escape in supernova remnants through gamma rays. MNRAS 490, 4317-4333.  https://doi.org/10.1093/mnras/stz2897.

\bibitem[\protect\citeauthoryear{Ellison \& Bykov}{2011}]{El11} Ellison, D. C. \& Bykov, A. M., 2011. Gamma-ray Emission of Accelerated Particles Escaping a Supernova Remnant in a Molecular Cloud. ApJ 731, 87.
https://doi.org/10.1088/0004-637X/731/2/87. 

\bibitem[\protect\citeauthoryear{Ergin et al.}{2017}]{Er17} Ergin, T., Sezer, A., Sano, H., Yamazaki, R., Fukui, Y., 2017. Recombining Plasma and Gamma-Ray Emission in the Mixed-morphology Supernova Remnant 3C 400.2. ApJ 842, 22.
https://doi.org/10.3847/1538-4357/aa72ee. 

\bibitem[\protect\citeauthoryear{Fujita et al.}{2009}]{Fu09} Fujita, Y., Ohira, Y., Tanaka, S. J., Takahara, F., 2009. Molecular Clouds as a Probe of Cosmic-Ray Acceleration in a Supernova Remnant. ApJ 707, L179-L183.
https://doi.org/10.1088/0004-637X/707/2/L179. 

\bibitem[\protect\citeauthoryear{Gabici et al.}{2009}]{Ga09} Gabici, S., Aharonian, F. A., Casanova, S., 2009. Broad-band non-thermal emission from molecular clouds illuminated by cosmic rays from nearby supernova remnants. MNRAS 396, 1629-1639.
https://doi.org/10.1111/j.1365-2966.2009.14832.x.

\bibitem[\protect\citeauthoryear{Kelner et al.}{2006}]{Ke06} Kelner, S. R., Aharonian, F. A., \& Bugayov, V. V., 2006. Energy spectra of gamma rays, electrons, and neutrinos produced at proton-proton interactions in the very high energy regime.  Phys. Rev. D 74, 034018. https://doi.org/10.1103/PhysRevD.74.034018.

\bibitem[\protect\citeauthoryear{Leahy \& Roger}{1991}]{LeRo91} Leahy, D. A., Roger, R. S., 1991. Radio Emission from the Supernova Remnant G160.9+2.6 (HB9). AJ 101, 1033.
https://doi.org/10.1086/115745.

\bibitem[\protect\citeauthoryear{Leahy \& Aschenbach}{1995}]{LeAs95} Leahy, D. A., Aschenbach, B., 1995. ROSAT X-ray observations of the supernova remnant HB 9. A\&A 293, 853-858.

\bibitem[\protect\citeauthoryear{Leahy \& Tian}{2007}]{LeTi07} Leahy, D. A., Tian, W. W., 2007. Radio spectrum and distance of the SNR HB9. A\&A 461, 1013-1018.
https://doi.org/10.1051/0004-6361:20065895.


\bibitem[\protect\citeauthoryear{Lee et al.}{2015}]{Le15} Lee, S.-H., Patnaude, D. J., Raymond, J. C., Nagataki, S., Slane, P. O., \& Ellison, D. C., 2015. Modeling Bright $\gamma$-Ray and Radio Emission at Fast Cloud Shocks. ApJ 806, 71.
https://doi.org/10.1088/0004-637X/806/1/71.

\bibitem[\protect\citeauthoryear{Li \& Chen}{2010}]{Li10} Li, H., Chen, Y., 2010. $\gamma$-rays from molecular clouds illuminated by accumulated diffusive protons from supernova remnant W28. MNRAS 409, L35-L38.
https://doi.org/10.1111/j.1745-3933.2010.00944.x. 

\bibitem[\protect\citeauthoryear{Malkov et al.}{2013}]{Ma13} Malkov, M. A., Diamond, P. H., Sagdeev, R. Z., Aharonian, F. A. \& Moskalenko, I. V., 2013. Analytic Solution for Self-regulated Collective Escape of Cosmic Rays from Their Acceleration Sites. ApJ 768, 73.
https://doi.org/10.1088/0004-637X/768/1/73. 

\bibitem[\protect\citeauthoryear{Mathis et al.}{1983}]{Ma83} Mathis, J. S., Mezger, P. G., \& Panagia, N., 1983. Interstellar radiation field and dust temperatures in the diffuse interstellar matter and in giant molecular clouds. A\&A 128, 212.

\bibitem[\protect\citeauthoryear{Ohira et al.}{2010}]{Oh10} Ohira, Y., Murase, K., Yamazaki, R., 2010. Escape-limited model of cosmic-ray acceleration revisited. A\&A 513, A17.
https://doi.org/10.1051/0004-6361/200913495.

\bibitem[\protect\citeauthoryear{Ohira et al.}{2011}]{Oh11} Ohira, Y., Murase, K., Yamazaki, R., 2011. Gamma-rays from molecular clouds illuminated by cosmic rays escaping from interacting supernova remnants. MNRAS 410, 1577-1582.
https://doi.org/10.1111/j.1365-2966.2010.17539.x.

\bibitem[\protect\citeauthoryear{Porter et al.}{2006}]{Po06} Porter, T. A., Mosalenko, I. V., \& Strong, A. W., 2006. Inverse Compton Emission from Galactic Supernova Remnants: Effect of the Interstellar Radiation Field. ApJ 648, L29-L32. https://doi.org/10.1086/507770.

\bibitem[\protect\citeauthoryear{Ptuskin \& Zirakashvili}{2003}]{Pu03} Ptuskin, V. S.\& Zirakashvili, V. N., 2003. Limits on diffusive shock acceleration in supernova remnants in the presence of cosmic-ray streaming instability and wave dissipation. A\&A 403, 1-10.
https://doi.org/10.1051/0004-6361:20030323.

\bibitem[\protect\citeauthoryear{Ptuskin \& Zirakashvili}{2005}]{Pu05} Ptuskin, V. S. \& Zirakashvili, V. N., 2005. On the spectrum of high-energy cosmic rays produced by supernova remnants in the presence of strong cosmic-ray streaming instability and wave dissipation. A\&A 429, 755-765.
https://doi.org/10.1051/0004-6361:20041517.

\bibitem[\protect\citeauthoryear{Rho \& Petre}{1998}]{RhPe98} Rho, J., Petre, R., 1998. Mixed-Morphology Supernova Remnants. ApJ 503, L167-L170.
https://doi.org/10.1086/311538.

\bibitem[\protect\citeauthoryear{Saito et al.}{2020}]{Sai20} Saito, M., Yamauchi, S., Nobukawa, K.K., Bamba, A., Pannuti, T.G., 2020. X-ray emission from the mixed-morphology supernova remnant HB 9. PASJ 72, 65.
https://doi.org/10.1093/pasj/psaa042. 

\bibitem[\protect\citeauthoryear{Sano et al.}{2017}]{Sa17} Sano, H. et al. 2017. Interstellar gas and X-rays toward the Young supernova remnant RCW 86; pursuit of the origin of the thermal and non-thermal X-ray. JHEAp 15, 1-18.
https://doi.org/10.1016/j.jheap.2017.04.002.

\bibitem[\protect\citeauthoryear{Sano et al.}{2019a}]{Sa19a} Sano, H. et al. 2019a. Possible Evidence for Cosmic-Ray Acceleration in the Type Ia SNR RCW 86: Spatial Correlation between TeV Gamma-Rays and Interstellar Atomic Protons. ApJ 876, 37.
https://doi.org/10.3847/1538-4357/ab108f. 

\bibitem[\protect\citeauthoryear{Sano et al.}{2019b}]{Sa19b} Sano, H. et al. 2019b. Discovery of Shocked Molecular Clouds Associated with the Shell-type Supernova Remnant RX J0046.5-7308 in the Small Magellanic Cloud. ApJ 881, 85.
https://doi.org/10.3847/1538-4357/ab2ade.

\bibitem[\protect\citeauthoryear{Sezer et al.}{2019}]{Se19} Sezer, A., Ergin, T., Yamazaki, R., Sano, H., Fukui, Y., 2019. Discovery of recombining plasma inside the extended gamma-ray supernova remnant HB9. MNRAS 489, 4300-4310.
https://doi.org/10.1093/mnras/stz2461.

\bibitem[\protect\citeauthoryear{Tang \& Chevalier}{2015}]{Ta15} Tang, X. \& Chevalier, R. A., 2015. Time-dependent Diffusive Shock Acceleration in Slow Supernova Remnant Shocks. ApJ 800, 103.
https://doi.org/10.1088/0004-637X/800/2/103.

\bibitem[\protect\citeauthoryear{Tashiro et al.}{2018}]{Tashiro2018}Tashiro, M. et al., 2018, in Society of Photo-Optical Instrumentation Engineers (SPIE) Conference Series, Vol. 10699, Proc. SPIE, 1069922.

\bibitem[\protect\citeauthoryear{The Fermi-LAT Collaboration}{2020}]{Fermi20} The Fermi-LAT Collaboration, 2020. Fermi Large Area Telescope Fourth Source Catalog. ApJS 247, 33.
https://doi.org/10.3847/1538-4365/ab6bcb.

\bibitem[\protect\citeauthoryear{Tuohy et al.}{1979}]{Tu79} Tuohy, I. R., Clark, D. H., Garmire, G. P., 1979. Observation of soft X-ray emission from the SNR HB 9. MNRAS 189, 59-63.
https://doi.org/10.1093/mnras/189.1.59P.

\bibitem[\protect\citeauthoryear{Yamauchi \& Koyama}{1993}]{Ya93} Yamauchi, S., Koyama, K., 1993. Hard X-Ray Emission from the Region near the Supernova Remnant HB 9 and the Radio Galaxy 4C 46.09. PASJ 45, 545-550.

\bibitem[\protect\citeauthoryear{Uchiyama et al.}{2010}]{Uc10} Uchiyama, Y., Blandford, R. D., Funk, S., Tajima, H., \& Tanaka, T., 2010. Gamma-ray Emission from Crushed Clouds in Supernova Remnants. ApJ 723, L122-L126.
https://doi.org/10.1088/2041-8205/723/1/L122.

\bibitem[\protect\citeauthoryear{Zhao et al.}{2020}]{Zh20} Zhao, H., Jiang, B., Li J., Chen, B., Yu, B., Wang, Y. 2020. A Systematic Study of the Dust of Galactic Supernova Remnants. I. The Distance and the Extinction. ApJ 891, 137.
https://doi.org/10.3847/1538-4357/ab75ef. 

\end{thebibliography}
\end{document}